\documentclass[prb,showpacs]{revtex4}
\usepackage{graphicx,psfrag,amssymb,amsmath}
\begin{document}
\title{Impurity effects in unconventional density waves in the unitary limit}
\author{Bal\'azs D\'ora}
\affiliation{The Abdus Salam ICTP, Strada Costiera 11, I-34014, Trieste, Italy}
\author{Attila Virosztek}
\affiliation{Department of Physics, Budapest University of Technology and 
Economics, H-1521 Budapest, Hungary}
\affiliation{Research Institute for Solid State Physics and Optics, P.O.Box
49, H-1525 Budapest, Hungary}
\author{Kazumi Maki}
\affiliation{Department of Physics and Astronomy, University of Southern
California, Los Angeles CA 90089-0484, USA}

\date{\today}

\begin{abstract}
We investigate the effect of strong, nonmagnetic impurities on quasi-one-dimensional 
conventional and unconventional density waves (DW and UDW). The conventional case remains unaffected similarly to s-wave 
superconductors in the presence of weak, nonmagnetic impurities. The thermodynamic properties of UDW were 
found to be identical to those of a d-wave superconductor in the unitary limit.
The real and imaginary part of the optical conductivity is determined for electric fields applied in the perpendicular directions.
A new structure can be present corresponding to excitations from the bound state at the Fermi energy to the gap maximum in 
addition to the usual peak at $2\Delta$. In the dc limit, universal electric conductivity is found.
\end{abstract}

\pacs{75.30.Fv, 71.45.Lr, 78.30.Jw, 72.15.Eb}

\maketitle

\section{Introduction}
The existence and behaviour of conventional (i.e. with constant gap) spin and charge density waves (SDW and CDW) 
is well documented\cite{gruner}. The thermodynamics of these systems was found to be very close to those of an
s-wave BCS superconductor due to the similar, fully gapped  density of states, but the transport properties
are completely different. After the discovery of unconventional superconductors, the extension of the field of density waves 
(DW)
into DW with wavevector dependent gap (termed unconventional) looks natural. In fact, after the earlier proposals
in the context of the excitonic insulator\cite{HR,kopaev}, this topic was rediscovered in the early 90's in various 
dimensions and systems\cite{Ners1,Ners2,Ners3,Schulz,GG,marston,Ozaki}. Since 
then, the 
realization
of unconventional or nodal density waves\cite{IO,Sudip} looks more and more possible: non-superconducting phase transitions without 
charge
or spin ordering have been detected in a number of materials, and one of the possible explanations is provided by the 
unconventional 
density 
wave scenario\cite{benfatto,castroneto,nagycikk,3dflux}. One of the main reasons of interest on UDW arises from high $T_c$ 
superconductors, where 
one of the competing models in the pseudogap phase is the d-density 
wave state.\cite{nayak,carbotte}.  

Recently we have studied the effect of impurities in the Born limit in unconventional density waves\cite{scatter}. This treatment
was justified from the fact, that this limit works very well for conventional density waves\cite{epl1}, and the investigated physical
quantities (for example the threshold electric field) showed convincing agreement with experimental data 
on $\alpha$-(BEDT-TTF)$_2$KHg(SCN)$_4$\cite{rapid,tesla,imperfect}. 
However, as is known
from high $T_c$ superconductors\cite{impurd-wave,hotta}, different impurities cause distinct effects on the same ground state: the Born 
and 
unitary
scattering limit seems to describe Ni and Zn impurities, respectively\cite{epl2}.
From this, it looks natural to extend our earlier analysis on the thermodynamic and transport properties to the unitary limit.

On the other hand, since conventional DW were mainly investigated in the Born scattering limit, it is instructive to study the effect
of unitary scatterers on this state, partly to complete the picture, and partly due the interesting physics of this subject.

In this paper we study impurity effects on quasi-one dimensional conventional and unconventional density waves at $T=0$. The basic 
advantage of quasi-one dimensionality is that the nesting condition can be fulfilled at arbitrary fillings. First we examine the effect 
of resonant scatterers on conventional density waves. Interestingly, the density of states and the thermodynamics remains unchanged
due to infinitely strong impurities, similarly to the effect of nonmagnetic impurities in s-wave 
superconductors in the Born limit\cite{parks,klasszikus}. This surprising
result follows from the fact that the nonmagnetic
impurity enhances the renormalised order parameter $\tilde\Delta_n$ in the unitary limit like it does in s-wave superconductor in 
the Born limit. As a result, a clean gap exists in the excitation spectrum for 
arbitrary impurity concentrations, and the low temperature physics is described by exponential functions with an activation energy. 
The unconventional situation gives more "conventional" results in the unitary limit. The thermodynamics looks very close to those
of a d-wave superconductor in the unitary limit\cite{impurd-wave,hotta}, and new localized states are visible around the Fermi 
energy. 
Similar phenomenon was 
observed in the density of states of isotropic p-wave superconductor\cite{pwave1,pwave2}, where a small island of states develops 
around 
the Fermi energy in 
the unitary limit. As a result of these new states at the Fermi energy, depending on the direction of the applied electric field and on 
the structure of the gap, new features are found for $\omega=\Delta$ in the optical spectra along with the pair breaking peak 
at $2\Delta$,
 where $\Delta$ is the gap maximum. 
In general the gapless nature of optical excitations was detected experimentally in $\alpha$-(BEDT-TTF)$_2$KHg(SCN)$_4$\cite{dressel}, 
which coincide with our theoretical results,  but for further conclusions more 
experiments are needed in the low temperature range.

\section{Formalism}

The single-particle electron thermal Green's function of DW is given by\cite{scatter,greendw}
\begin{equation}
G^{-1}({\bf k}, i\omega_n)=i\omega_n-\xi({\bf k})\rho_3-\rho_1\sigma_3\textmd{Re}\Delta({\bf 
k})-\rho_2\sigma_3\textmd{Im}\Delta({\bf k}),\label{Green0}
\end{equation}
where $\rho_i$ and $\sigma_i$ ($i=1,2,3$) are the Pauli matrices acting on momentum and spin space, respectively, and for (U)CDW 
$\sigma_3$ should be replaced by $1$. $\Delta({\bf k})=\Delta e^{i\phi}f({\bf k})$, $f({\bf k})=1$ in the conventional case and
$\cos(bk_y)$ or $\sin(bk_y)$ in the unconventional case. $\phi$ is the unrestricted phase (due to incommensurability) of the density 
wave, $\xi(\bf k)$ is the kinetic energy spectrum, $\omega_n$ is the fermionic Matsubara frequency.
%Since in the present analysis we are not interested in phenomena related to the phase of the DW (i.e. sliding), we can fix it to 
%$\phi=0$ for simplicity. 

The Hamiltonian describing the interaction of the electrons with nonmagnetic impurities is given by 
\begin{eqnarray}
H_1=\frac1V \sum_{{\bf k,q}, \sigma, j}e^{-i\bf{qR}_j}\Psi_\sigma^+({\bf k+q})U({\bf R}_j)
\Psi_\sigma({\bf k}),\\
U({\bf R}_j)=\left( \begin{array}{cc}
         U(0) & U({\bf Q})e^{-i{\bf QR}_j} \\
         \overline{U({\bf Q})}e^{i{\bf QR}_j} & U(0)
         \end{array}
 \right),\label{impurity}
\end{eqnarray}
${\bf R}_j$ is the position of the $j$-th impurity atom, $\bf Q$ is the nesting vector.

The explicit wavevector dependence of the matrix
elements\cite{haran,rapid} is neglected since no important changes are expected from it.
Following the method of Ref. \onlinecite{scatter}, the self energy correction from impurities is given by
\begin{gather}
\Sigma_{\bf R}(i\omega_n)=n_i\left(U({\bf R})^{-1}-\int\frac{d^3p}{{2\pi}^3}G({\bf p},i\omega_n)\right)^{-1}
\end{gather}
where the the $\bf R$ index in $\Sigma_{\bf R}(i\omega_n)$ means the position of an 
impurity over which the average should be taken, $n_i$ is
the impurity concentration. Here following the standard approach only noncrossing diagrams were taken into account. By fixing the 
ratio of $U({\bf Q})/U(0)$ and taking $U(0)$ to infinity, 
the self energy is given by
\begin{equation}
\Sigma(i\omega_n)=-n_i\left(\int\frac{d^3p}{{2\pi}^3}G({\bf p},i\omega_n)\right)^{-1}
\label{selfenergy}
\end{equation}
We note here that in the special case of $U(0)=|U({\bf Q})|$, the $U(\bf R)$ matrix is singular, and the above calculations are not 
valid, but this condition corresponds to the fact that 
 in real space the electron-impurity interaction is ultra short range, namely $U({\bf r})\sim\delta({\bf r})$, which is not the 
case in real systems.  
From Eq. (\ref{selfenergy}), the self energy correction in the conventional case is obtained as
\begin{gather}
\Sigma(i\omega_n)=-\frac{\Gamma}{\sqrt{u_n^2+1}}\left(\begin{array}{cc}
         iu_n & -e^{i\phi} \\
         -e^{-i\phi} & iu_n
         \end{array}
 \right)
\end{gather}
where $g(0)$ is the density of states per spin in the normal state at the Fermi energy, $\Gamma=2n_i/\pi g(0)$, 
$u_n=\tilde\omega_n/\tilde\Delta_n$, 
$\tilde\omega_n$ 
and $\tilde\Delta_n$ are the renormalized frequency and gap:
\begin{eqnarray}
\omega_n=\tilde\omega_n\left(1-\frac{\Gamma}{\sqrt{\tilde\omega_n^2+\tilde\Delta_n^2}}\right),\label{conv1}\\
\Delta=\tilde\Delta_n\left(1-\frac{\Gamma}{\sqrt{\tilde\omega_n^2+\tilde\Delta_n^2}}\right)\label{conv2}.
\end{eqnarray}
From this, the relation $u_n=\tilde\omega_n/\tilde\Delta_n=\omega_n/\Delta$ holds.
In the unconventional case, the self energy correction is obtained as
\begin{equation}
\Sigma(i\omega_n)=\Gamma \frac{\pi}{2}\dfrac{\sqrt{u_n^2+1}}{u_n K\left(\frac{1}{\sqrt{u_n^2+1}}\right)},
\end{equation}
where $K(z)$ is the complete elliptic integral of the first kind. The gap remains unrenormalized due to the zero average of the gap 
function $f({\bf k})$ over the 
Fermi surface, $u_n=\tilde\omega_n/\Delta$, and the Matsubara frequency is renormalized as 
\begin{equation}
\omega_n=\Delta u_n-\Gamma\frac{\pi}{2}\dfrac{\sqrt{u_n^2+1}}{u_n K\left(\frac{1}{\sqrt{u_n^2+1}}\right)}.
\end{equation}
This is the same as in d-wave superconductors, the presence of backscattering ($U({\bf Q})$) drops out from the calculation and does 
not modify the result as in the Born limit. It is useful to introduce the quantity $u_n(\omega_n=0)=C_0$, which is determined from
\begin{equation}
K\left(\frac{1}{\sqrt{C_0^2+1}}\right)=\frac{\pi\Gamma}{2\Delta}\frac{\sqrt{1+C_0^2}}{C_0^2},
\end{equation}
and will be used in further calculations.

\section{Conventional density wave}

The density of states (DOS) is obtained as
\begin{equation}
\frac{N(\omega)}{g(0)}=\textmd{Im}\frac{u}{\sqrt{1-u^2}}=\frac{|\omega|}{\sqrt{\omega^2-\Delta^2}}\Theta(|\omega|-2\Delta),
\end{equation}
where $u=iu_n(i\omega_n=\omega+i\delta)$, $\Theta(z)$ is the Heaviside function and the second equality follows from Eqs. (\ref{conv1}) 
and (\ref{conv2}). Hence the density of states remains unchanged in the presence of infinitely strong impurities, which is identical to 
the behaviour of s-wave superconductors in the presence of weak nonmagnetic impurities\cite{klasszikus,parks}. As a result of the 
unchanged, gapped density of 
states, the thermodynamic properties such as the transition temperature or the specific heat remain the same as in the pure 
conventional density wave. This can be understood from the simple one impurity picture studied by T\"utt\H o and Zawadowski in Ref. 
\onlinecite{tutto1,tutto2}.
The basic effect of impurities, the pinning comes from the interference between the Friedel oscillation and the density wave. For
infinitely strong backscattering, however, the phase of outgoing electron is the opposite of the incoming one, hence the Friedel 
oscillation dies 
out\cite{tutto1}. This simple picture has to be modified in the presence of the DW condensate, but the lack of Friedel oscillation 
still holds in 
the unitary limit, and no pinning is possible.

In the transport properties, there are differences between the pure and impure systems. The optical conductivity for electric fields
perpendicular to the chain direction still exhibits a clean gap for $\omega<2\Delta$, but the divergent peak at $2\Delta$ turns into a 
sharp but finite cusp.

\section{Thermodynamics of unconventional density waves}

\begin{figure}[t!]
%\hspace*{-2cm}
\psfrag{x}[t][b][1.2][0]{$\Gamma/\Gamma_c$}
\psfrag{y}[b][t][1.2][0]{$\Delta(0,\Gamma)/\Delta_{00}$, $T_c/T_{c0}$ and
$N(0,\Gamma)/g(0)$}
{\includegraphics[width=7cm,height=7cm]{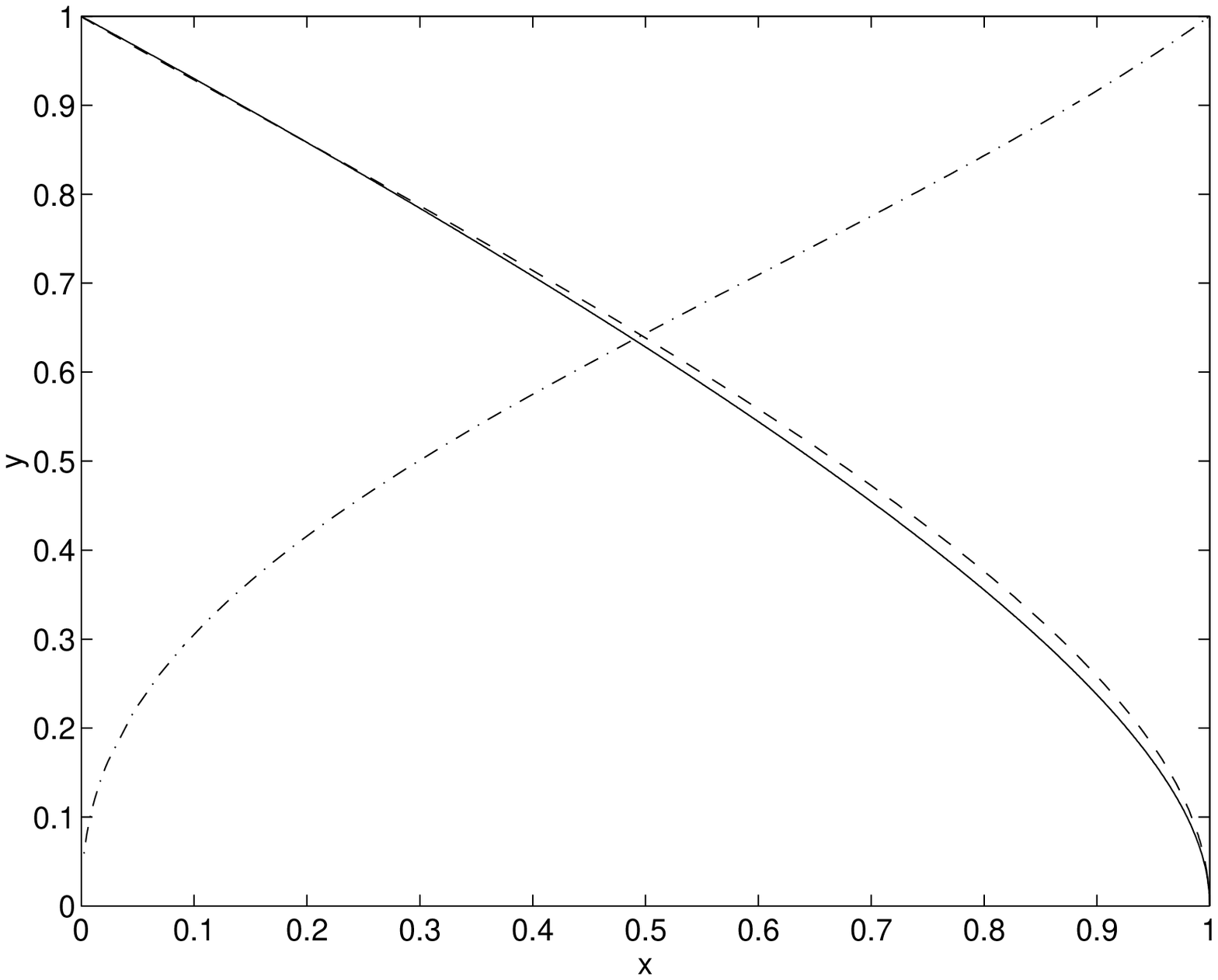}}
\hspace*{1cm}
\psfrag{x}[t][b][1.2][0]{$\omega/\Delta$}
\psfrag{y}[b][t][1.2][0]{$N(\omega)/g(0)$}
\includegraphics[width=7cm,height=7cm]{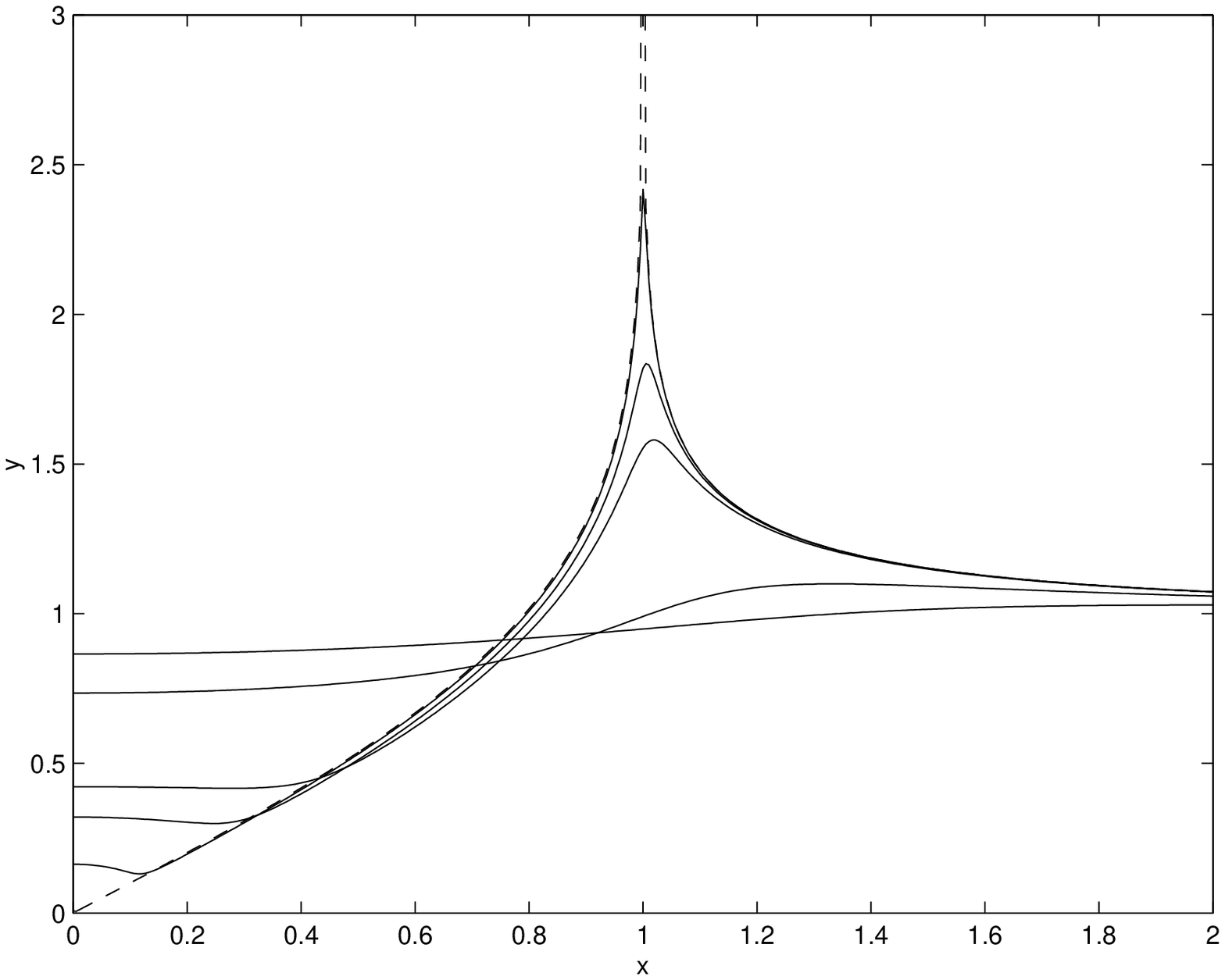}
\psfrag{xx}[t][b][0.8][0]{$\omega/\Delta$}
\psfrag{yy}[b][t][0.8][0]{$N(\omega)/g(0)$}
\psfrag{0}[][b][0.5][0]{0}
\psfrag{0.1}[t][b][0.5][0]{0.1}
\psfrag{0.2}[t][b][0.5][0]{0.2}
\psfrag{1}[][][0.5][0]{0 }
\psfrag{1.1}[][][0.5][0]{0.1 }
\psfrag{1.2}[][][0.5][0]{0.2 }

\vspace*{-6.7cm}\hspace*{12cm}
\includegraphics[width=2.6cm,height=2.7cm]{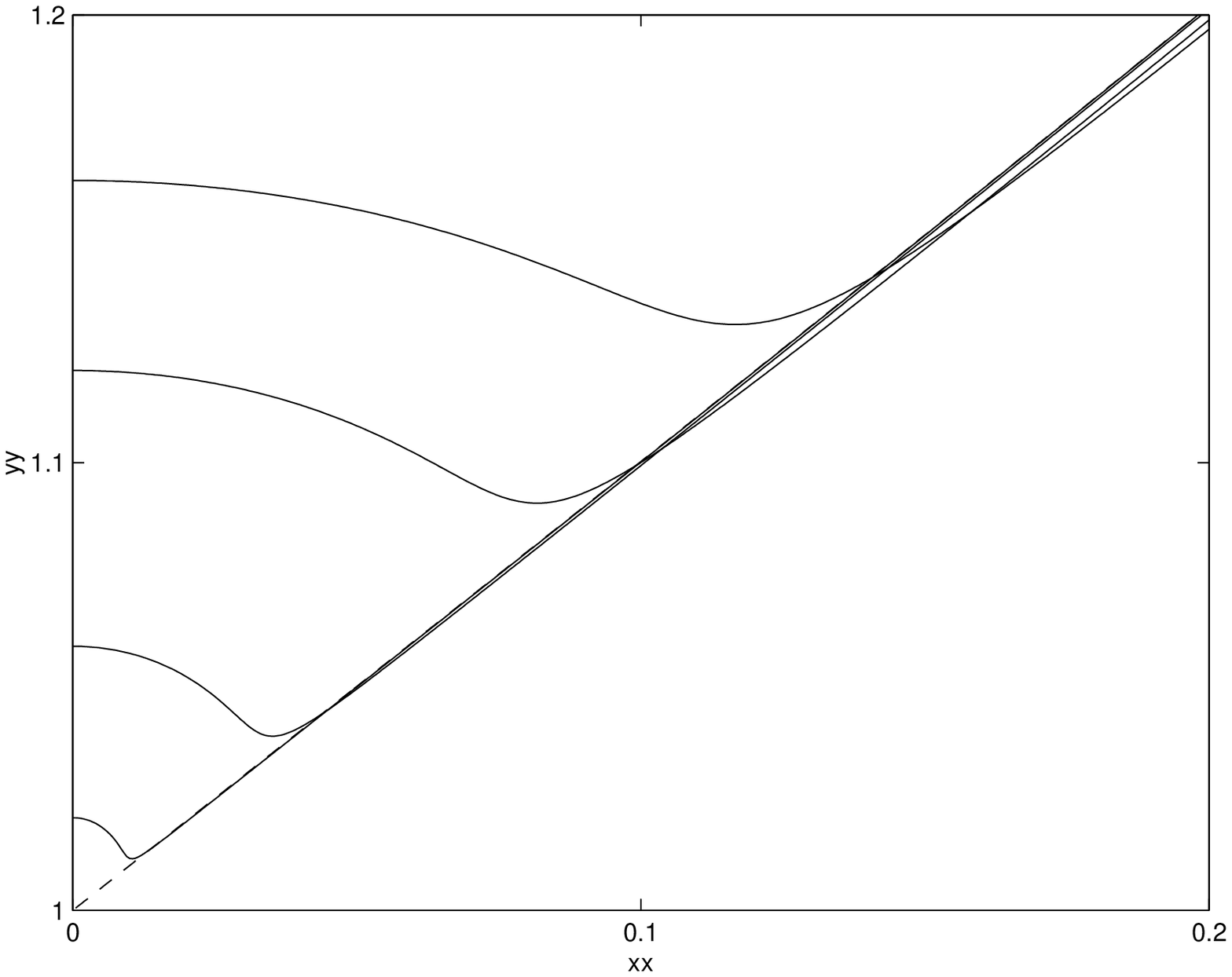}

\vspace*{5cm}
\caption{In the left panel, $\Delta(0,\Gamma)/\Delta_{00}$ (dashed line), $T_c/T_{c0}$ (solid
line) and $N(0,\Gamma)/g(0)$ (dashed-dotted line) are shown as a function of
$\Gamma/\Gamma_c$. In the right panel, the density of states is shown for $\Gamma/\Delta=0$ (dashed 
line), $0.01$, $0.05$, $0.1$, $0.5$ and $1$ with increasing 
$N(0)$. The inset 
shows the localized state around the Fermi energy for $\Gamma/\Delta=0$ (dashed line), $0.0001$, $0.001$, $0.005$ and $0.01$ with 
increasing $N(0)$.}
\label{fig:dos}
\end{figure}

The density of states is obtained as
\begin{equation}
\frac{N(\omega)}{g(0)}=\frac{2}{\pi}\textmd{Im}\frac{u}{\sqrt{1-u^2}} 
K\left(\frac{1}{\sqrt{1-u^2}}\right)=\textmd{Im}\frac{\Gamma}{\omega-\Delta u},
\end{equation}
where $u=iu_n(i\omega_n=\omega+i\delta)$.
It is identical to those of a d-wave superconductor in the presence of nonmagnetic impurities in the unitary limit\cite{hottacomment}, 
and so does the
thermodynamics as well, which can be borrowed from d-wave superconductors (Refs. \onlinecite{impurd-wave,hotta} and references 
therein). Consequently the change of the 
transition temperature is 
given by the Ab\-ri\-kos\-ov-Gor'kov formula:
\begin{equation}
-\ln\left(\frac{T_c}{T_{c_0}}\right)=\psi\left(\frac12+\rho\right)-
\psi\left(\frac12
\right),
\end{equation}
where $T_c$ and $T_{c_0}$ are the transition temperature of the impure
and clean
system, respectively, $\rho=\Gamma/2\pi T_c$, $\psi(z)$ is the
digamma function. This formula holds also in the Born scattering limit\cite{scatter} for both conventional and unconventional density 
waves as well 
as for unconventional superconductors in the presence 
of impurities
considered either in Born or in resonant scattering limit\cite{szupravezetes}.
The critical impurity scattering rate is obtained as
\begin{equation}
\Gamma_c=\frac{\pi T_{c_0}}{2\gamma}=\frac{\sqrt e \Delta_{00}}{4}.
\end{equation}
Using the parameters of $\alpha$-(BEDT-TTF)$_2$KHg(SCN)$_4$, namely $T_c=10$K, $v_F=6\times 10^4$m/s and lattice constant in the chain 
direction $a=10^{-9}$m, the critical concentration is estimated as $n_i=0.001$.  
In Fig. \ref{fig:dos}, we show the transition temperature, the residual density of states (i.e. $N(0)$) and the zero temperature 
gap 
coefficient as a function of the scattering rate. 
The density of states exhibits localized state due to impurities around the Fermi energy superimposed on the usual gapless 
density 
of states of the pure system, which is manifested in the nonmonotonic nature 
of the DOS close to the Fermi energy\cite{hottacomment}, as it is shown in Fig. \ref{fig:dos}. This state gives rise to a feature
 in the optical 
response as we will demonstrate it 
later. The identification of the localized or bound states is clearer for fully gapped systems: similar localized states were 
found in conventional density waves in the presence of one single, relatively strong impurity\cite{tutto2}, and also the small island 
of 
states in 
unitary isotropic p-wave superconductors\cite{pwave1,pwave2} around 
the Fermi energy signals the 
presence of new bound states.

\section{Optical conductivity}

\begin{figure}[h!]
\psfrag{x}[t][b][1.2][0]{$\omega/\Delta_{00}$}
\psfrag{y}[b][t][1.2][0]{Re$\sigma_{yy}^{cos}(\omega)2\Delta_{00}/e^2g(0)v_y^2$}
{\includegraphics[width=7cm,height=7cm]{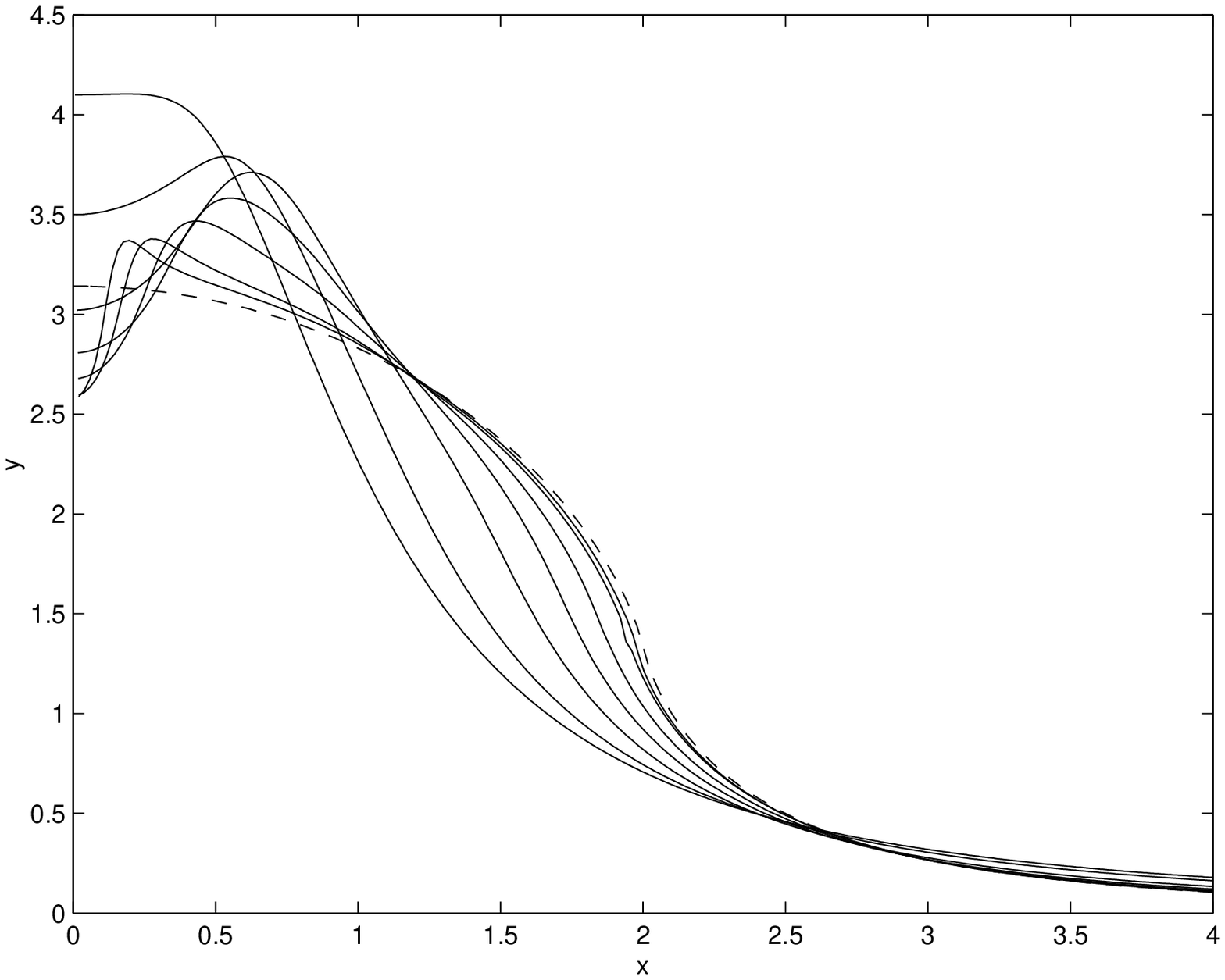}}
\hspace*{1cm}
\psfrag{x}[t][b][1.2][0]{$\omega/\Delta_{00}$}
\psfrag{y}[b][t][1.2][0]{$\omega$Im$\sigma_{yy}^{cos}(\omega)/2 e^2g(0)v_y^2$}
{\includegraphics[width=7cm,height=7cm]{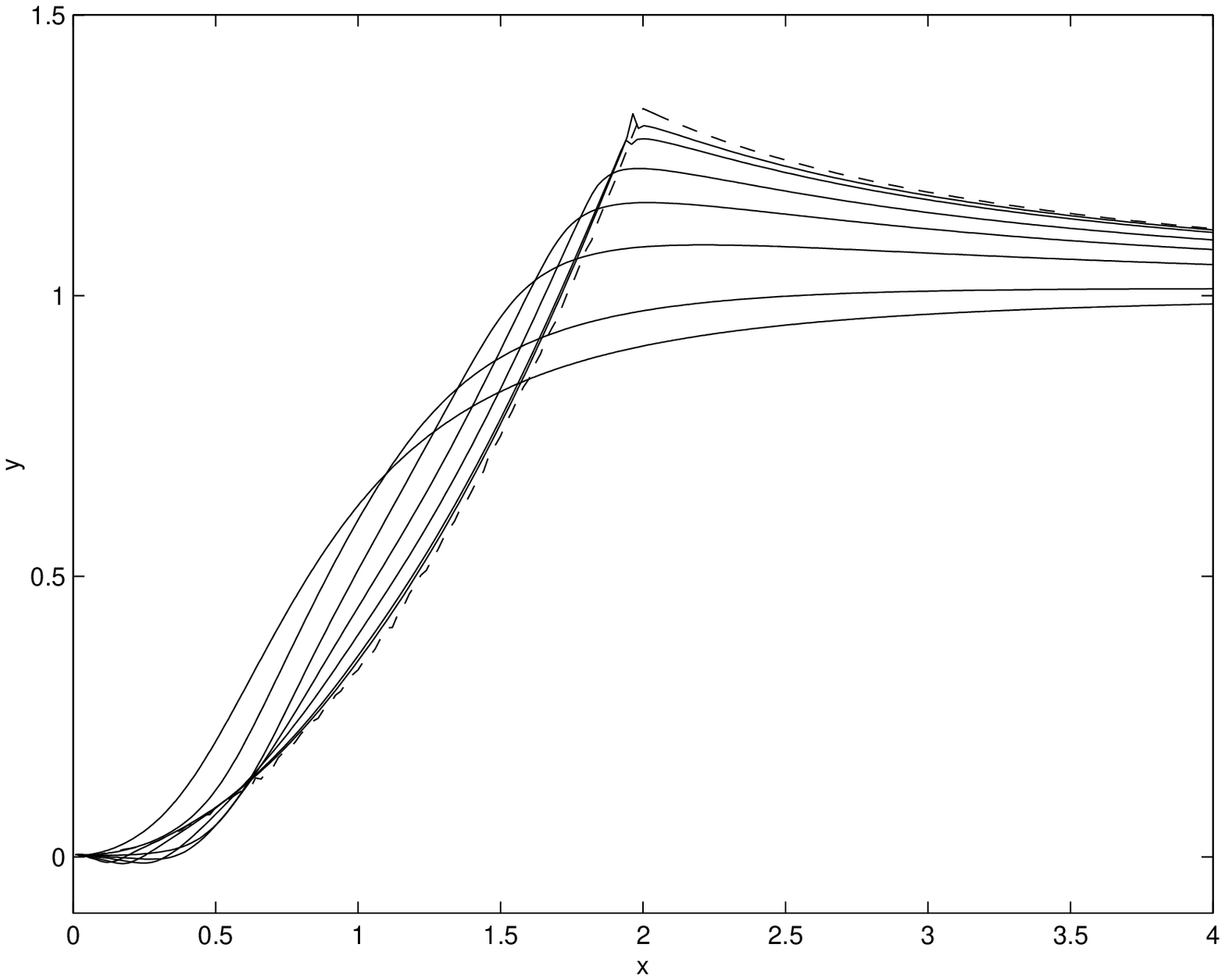}}

\caption{Real and imaginary part of the electric conductivity in the $y$ direction for
$\Delta({\bf k})=\Delta\cos(bk_y)$ are plotted as a function of the reduced energy for
different scattering amplitudes: $\Gamma/\Delta=0$ (dashed line), $0.01$, $0.02$, $0.05$, $0.1$, $0.2$, $0.5$ and $1$ with  
 decreasing Re$\sigma(2\Delta)$, Im$\sigma(2\Delta)$.
\label{fig:uvyyc}}
\end{figure}

\begin{figure}[h!]
\psfrag{x}[t][b][1.2][0]{$\omega/\Delta_{00}$}
\psfrag{y}[b][t][1.2][0]{Re$\sigma_{yy}^{sin}(\omega)2\Delta_{00}/e^2g(0)v_y^2$}
{\includegraphics[width=7cm,height=7cm]{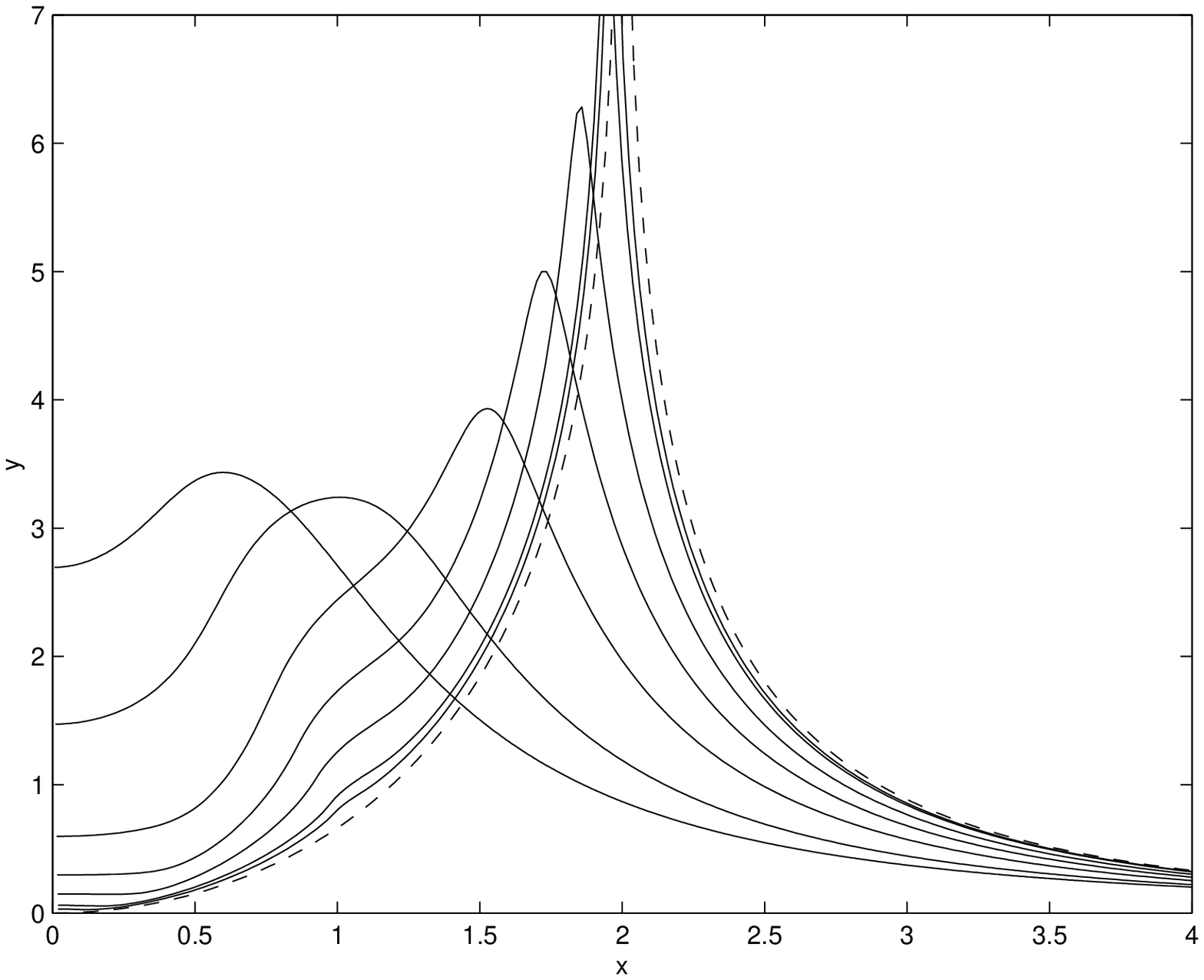}}
\hspace*{1cm}
\psfrag{x}[t][b][1.2][0]{$\omega/\Delta_{00}$}
\psfrag{y}[b][t][1.2][0]{$\omega$Im$\sigma_{yy}^{sin}(\omega)/2 e^2g(0)v_y^2$}
{\includegraphics[width=7cm,height=7cm]{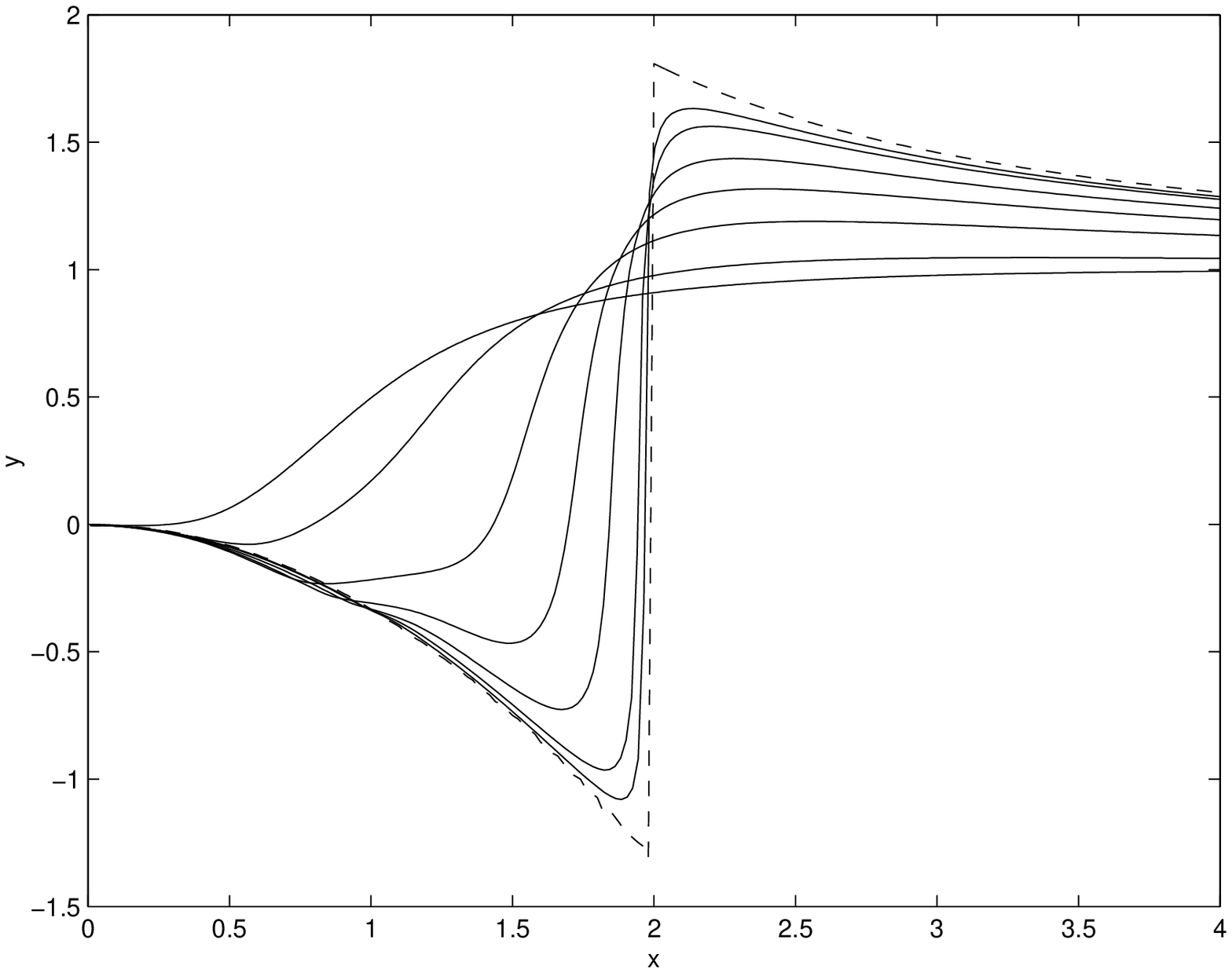}}

\caption{Real and imaginary part of the electric conductivity in the $y$ direction for
$\Delta({\bf k})=\Delta\sin(bk_y)$ are plotted as a function of the reduced energy for
different scattering amplitudes: $\Gamma/\Delta=0$ (dashed line), $0.01$, $0.02$, $0.05$, $0.1$, $0.2$, $0.5$ and $1$ 
with decreasing Re$\sigma(2\Delta)$, increasing Im$\sigma(\Delta)$.
\label{fig:uvyys}}
\end{figure}

We calculate the optical conductivity for electric fields perpendicular to the conducting chain. In this case collective modes do 
not show up or can be neglected, depending on the explicit wavevector dependence of the gap\cite{rpa}. Henceforth the optical response 
is 
calculated from the one bubble contribution, where self energy and vertex correction are taken into account in the noncrossing 
approximation.
The real and imaginary part of the optical conductivity at $T=0$ is given by\cite{epl1,epl2}:
\begin{gather}
\textmd{Re}\sigma_{aa}(\omega)=\frac{e^2 g(0)v_a^2}{\omega}\frac{4}{\Delta \pi}\textmd{Re}I(\omega),\\
\omega\textmd{Im}\sigma_{aa}(\omega)=e^2 g(0)v_a^2\frac{4}{\Delta \pi}\left(\textmd{Im}I(\omega)+2\int_{0}^\infty 
\textmd{Im}F(u(x),u(x+\omega))dx\right),
\end{gather}
where
\begin{equation}
I(\omega)=\int_{0}^\omega (F(u(\omega-x),\overline{u(-x)})-F(u(\omega-x),u(-x)))dx
\end{equation}
and $v_x=v_F$, $v_y=\sqrt 2 bt_b$ and $v_z=\sqrt 2 ct_c$.
In the following we discuss the different cases depending on the electric field orientation and on the gap.

i.\hspace*{8mm} $\Delta({\bf k})=\Delta\cos(k_yb)$, $a=y$:
\begin{gather}
F(u,{u'})=\frac{1}{{u'}^2-u^2}\left[\sqrt{1-{u'}^2}\left(E'\left(-u{u'}-\frac23+\frac{{u'}^2}{3}\right)+
K'\left(u{u'}-\frac{{u'}^2}{3}\right)\right)\right.+\nonumber
\\
\left.+\sqrt{1-u^2}\left(E\left(u{u'}+\frac23-\frac{u^2}{3}\right)+K\left(-u{u'}+\frac{u^2}{3}\right)\right)\right].
\end{gather}
In the definition of the different $F(u,u')$ functions the argument of $E$
and $K$ is $1/\sqrt{1-u^2}$, while for $E'$ and $K'$ $1/\sqrt{1-{u'}^2}$
has to be used. In the present case, vertex corrections vanish similarly to the Born limit
due to
the mismatch of wavevector dependence of the velocity and the gap. In the real part a small peak develops close to $\omega=0$, and 
moves to higher 
frequencies with increasing impurity concentration, but finally disappears as curves take more and more the
form of a Lorentzian. Here the presence of bound states cannot be seen because the weight of scattering from the Fermi energy to the 
gap maximum is zero due to the zero velocity of quasiparticles at the latter point. In the imaginary part the cusp at 
$\omega=2\Delta$ smoothens as $\Gamma$ increases, as it is seen in Fig. \ref{fig:uvyyc}. The dc conductivity is calculated at $T=0$ as:
\begin{equation}
\sigma_{yy}^{dc,cos}=e^2g(0)v_y^2\frac{4}{\Delta\pi}\left(E\sqrt{1+C_0^2}
-\frac{\pi \Gamma}{2\Delta}\right).
\end{equation}
In the dc conductivities, the argument of $E$ and $K$ is $1/\sqrt{1+C_0^2}$.
\vspace{2mm}

ii.\hspace*{8mm} $\Delta({\bf k})=\Delta\sin(k_yb)$, $a=y$:
\begin{gather}
F(u,{u'})=\frac{1}{{u'}^2-u^2}\left[\sqrt{1-u^2}E\left(-u{u'}+\frac43+\frac{u^2}{3}\right)
-\sqrt{1-{u'}^2}E'\left(-u{u'}+\frac43+\frac{{u'}^2}{3}\right)-\right.\nonumber \\
\left.
-\frac{{u'}^2}{\sqrt{1-{u'}^2}}K'\left(-u{u'}+\frac23+\frac{{u'}^2}{3}\right)+\frac{u^2}{\sqrt{1-u^2}}K\left(-u{u'}+\frac23+\frac{u^2}{3}\right)\right]+\nonumber
\\
+\dfrac{\Gamma\pi\sqrt{1-u^2}\sqrt{1-{u'}^2}}{2\Delta u u' K K'}\dfrac{1}{(u+{u'})^2}\dfrac{\left(E'\sqrt{1-{u'}^2}-E\sqrt{1-u^2}+
\dfrac{{u'}^2}{\sqrt{1-{u'}^2}}K'-\dfrac{u^2}{\sqrt{1-u^2}}K\right)^2}{1+\dfrac{\Gamma\pi}{2\Delta}
\dfrac{1}{u+{u'}}\left(\dfrac{\sqrt{1-{u'}^2}}{u'K'}+\dfrac{\sqrt{1-u^2}}{uK}\right)}.
\end{gather}
The real part of the conductivity exhibits a sharp peak at $2\Delta$ and a small bump at $\Delta$, indicating excitations from the 
localized state to the gap maximum for low concentrations. By increasing 
$\Gamma$, the former is suppressed and the latter becomes dominant. The imaginary part changes sign sharply at $2\Delta$, and a dip is 
present at $\Delta$, as can be readily seen in Fig. \ref{fig:uvyys}. 
The dc conductivity is obtained as:
\begin{equation}
\sigma_{yy}^{dc,sin}=e^2g(0)v_y^2\frac{2}{\Delta}\frac{C_0^2(K-E)}{\pi\sqrt{1+C_0^2}-\Delta C_0^2E/\Gamma}.
\end{equation}
The conductivity is shown in Fig. \ref{fig:uvyys}. 
\vspace{2mm}

iii.\hspace*{8mm} $\Delta({\bf k})=\Delta\sin(k_yb)$ or $\Delta\cos(k_yb)$, $a=z$:
\begin{gather}
F(u,{u'})=\frac{1}{2({u'}^2-u^2)}\left(2\sqrt{1-u^2}E-2\sqrt{1-{u'}^2}E'
+K'\frac{{u'}(u-{u'})}{\sqrt{1-{u'}^2}}+K\frac{u(u-{u'})}{\sqrt{1-u^2}}\right),
\end{gather}
the vertex corrections vanish because the velocity depends on different perpendicular
wavevector component ($k_z$) than the gap ($k_y$). As $\Gamma$ increases, the dominance of the $\Delta$ peak becomes more prominent 
than in the previous case in the real part of the conductivity. The imaginary part of the conductivity is zero for $\omega<2\Delta$ in 
the pure case, and exhibits  a sharp 
peak at $2\Delta$. 
The dc conductivity is obtained at $T=0$ as
\begin{equation}
\sigma_{zz}^{dc}=2e^2g(0)v_z^2\frac{E}{\Delta\pi\sqrt{C_0^2+1}},
\end{equation} 
\begin{figure}[h!]
\psfrag{x}[t][b][1.2][0]{$\omega/\Delta_{00}$}
\psfrag{y}[b][t][1.2][0]{Re$\sigma_{zz}(\omega)\Delta_{00}/e^2g(0)v_y^2$}
{\includegraphics[width=7cm,height=7cm]{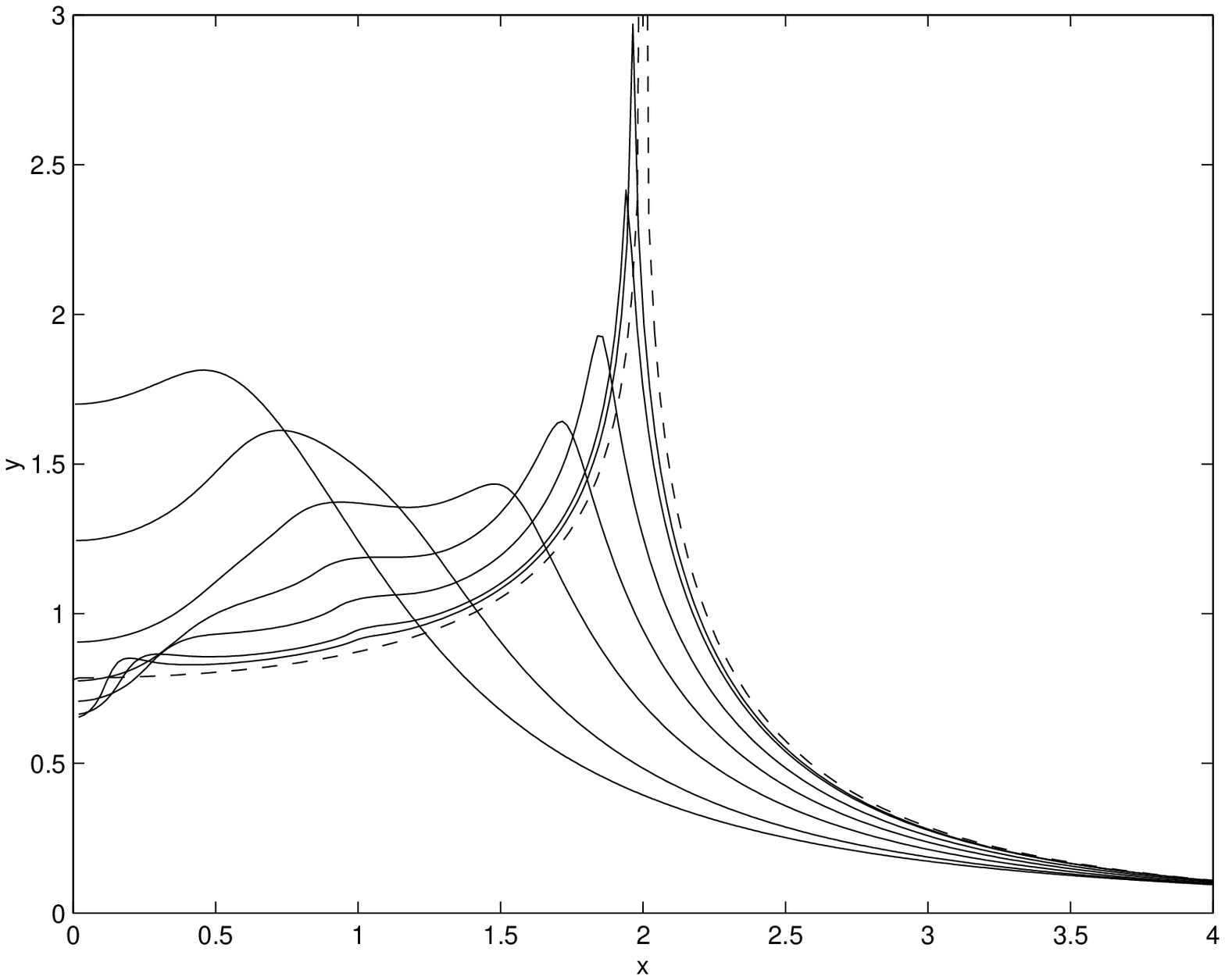}}
\hspace*{1cm}
\psfrag{x}[t][b][1.2][0]{$\omega/\Delta_{00}$}
\psfrag{y}[b][t][1.2][0]{$\omega$Im$\sigma_{zz}(\omega)/2 e^2g(0)v_y^2$}
{\includegraphics[width=7cm,height=7cm]{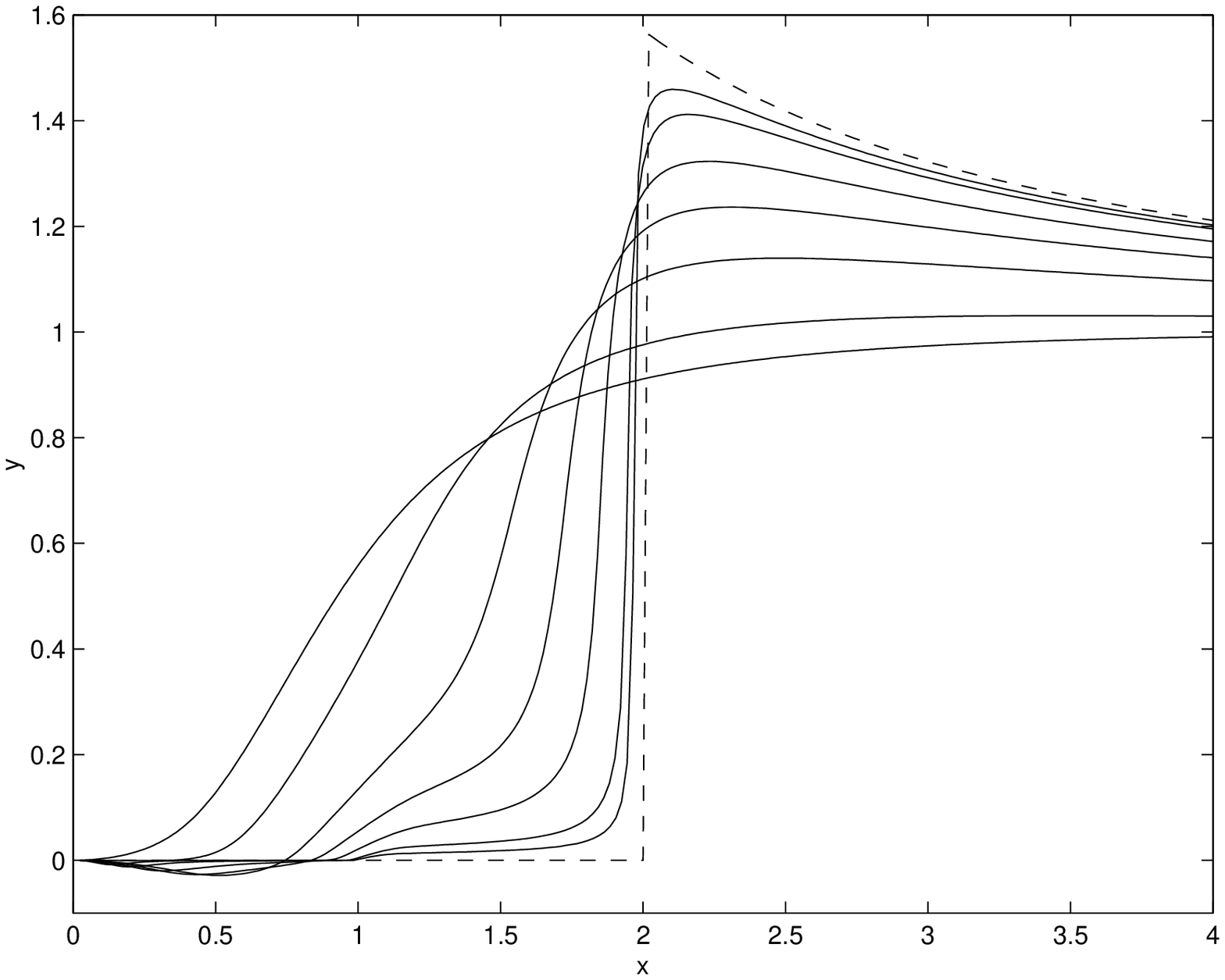}}

\caption{Real and imaginary part of the electric conductivity in the $z$ direction are plotted as a function of the reduced energy 
for
different scattering amplitudes: $\Gamma/\Delta=0$ (dashed line), $0.01$, $0.02$, $0.05$, $0.1$, $0.2$, $0.5$ and $1$ with
decreasing Re$\sigma(2\Delta)$, increasing Im$\sigma(\Delta)$.
\label{fig:uvzz}}
\end{figure}
\vspace{2mm}
The latter two cases seems to be consistent with experimental data on $\alpha$-(BEDT-TTF)$_2$KHg(SCN)$_4$\cite{dressel} as far as the 
gapless nature 
of the optical response is considered, while the former with almost monotonically decreasing Re$\sigma(\omega)$ is different 
from the measured data.
We refrain here from the evaluation of quasiparticle part of in chain conductivity, because the sliding collective mode
associated with the phase of the condensate dominates this response\cite{rpa}. We note however that the 
quasiparticle part of $\sigma_{xx}(\omega)$ is expected to behave very 
similarly to $\sigma_{zz}(\omega)$.

\begin{figure}[h!]
\psfrag{x}[t][b][1.2][0]{$\Gamma/\Gamma_c$}
\psfrag{y}[b][t][1.2][0]{$\sigma(0)\Delta_{00}/e^2 g(0) v_{y,z}^2$}
\centering{\includegraphics[width=7cm,height=7cm]{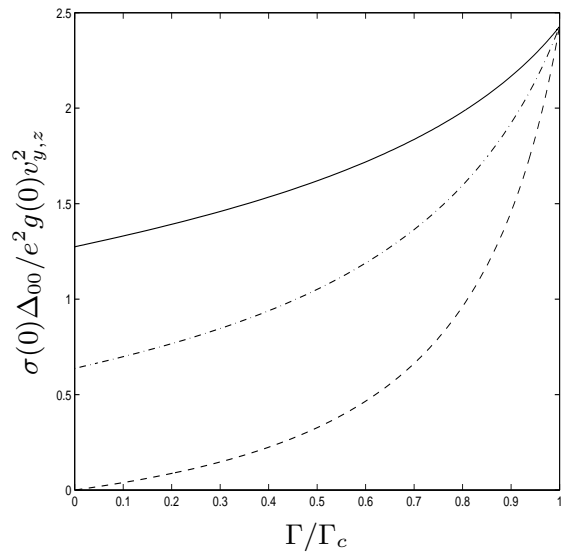}}
\caption{The dc conductivity is plotted at $T=0$ as a function of the reduced
scattering rate for a cosinusoidal (sinusoidal)
gap in the $y$ direction: solid (dashed line) and in the $z$ direction:
dashed-dotted line. \label{fig:dc}}
\end{figure}

The dc conductivities are shown in Fig. \ref{fig:dc} at $T=0$ as a function
of the impurity scattering parameter. In the perpendicular direction, the dc
conductivities take the same value at the critical scattering parameter, namely $e^2g(0)v_{y,z}^2/\Gamma_c$. 
Surprisingly, for small concentrations the dc response increases linearly with $\Gamma$, as opposed to the almost $\Gamma$ independent 
behaviour in 
the Born limit\cite{scatter}. This increasing behaviour is attributed to the fact that the creation of zero energy quasiparticles due 
to impurities is more 
efficient than the scattering of quasiparticles by impurities\cite{universal}.
It is worth mentioning that in the dc conductivity the $\Gamma\rightarrow 0$ and $\omega\rightarrow 
0$ limit cannot be exchanged, as it is seen in Figs. \ref{fig:uvyyc}, \ref{fig:uvyys}, \ref{fig:uvzz} and \ref{fig:dc}, this is why we 
obtain different dc conductivities in the pure case depending in the order of limits.
However, we believe the right procedure is shown in Fig. \ref{fig:dc}, where the $\omega\rightarrow 0$ limit is taken first. 
The dc conductivity in all cases turns out to be universal\cite{lee}, since regardless to the scattering limit it 
takes the 
same value as $\Gamma\rightarrow 0$, namely $\sigma^{dc,cos}_{yy}=e^2g(0)v_y^2 4/\Delta_{00}\pi$, $\sigma^{dc}_{zz}=e^2g(0)v_z^2 
2/\Delta_{00}\pi$ and $\sigma^{dc,sin}_{yy}=0$. The last equality holds since the electric current operator 
vanishes on the nodal points of the gap.

\section{Conclusion}

We have studied the effect of nonmagnetic impurities in conventional and unconventional 
density waves in the unitary scattering limit in the standard noncrossing approximation. In the conventional case, no changes are found 
in the thermodynamics compared to the 
pure system, similarly to s-wave superconductors in the Born limit. In the presence of one single, infinitely strong impurity
the Friedel oscillation disappears\cite{tutto1,tutto2}, since the phase of the incoming and outgoing electron is opposite. Consequently 
there is no interference between the density wave and the Friedel oscillation, and there is no pinning. In the presence of impurities 
with finite concentration, this simple picture seems to survive and the effect of impurities is canceled from the thermodynamics.

As opposed to this, in the unconventional case, the thermodynamic properties are identical to those of a d-wave superconductor in the 
unitary limit. From the density of states, it is obvious that electrons are localized close to the Fermi energy, while at larger 
energies they remain almost unaffected by the presence of impurities (aside from the broadening of the $\omega=\Delta$ peak).
Also the change in the transition temperature is given by the Abrikosov-Gor'kov formula, which was also found to be  valid in the Born 
limit\cite{scatter}.
Both the real and imaginary part of the optical conductivity seems to reflect the presence of localized states around the Fermi energy 
at certain gap structures by displaying a new bump at $\omega=\Delta$. This new feature seems to dominate over the pair 
breaking $\omega=2\Delta$ peak as the impurity concentration increases. 
We found universal electric conduction\cite{lee} in the dc limit.
The comparison of the optical conductivity with experimental data seems to be difficult due to the lack of 
consistent investigations. This can be attributed to the fact, that the material which possesses most likely quasi-one dimensional UCDW 
ground state, the $\alpha$-(BEDT-TTF)$_2$KHg(SCN)$_4$ salt enters into its new phase at $10$K, and optical experiments below this 
temperature are very difficult. The only available data\cite{dressel} reports some kind of pseudogap behaviour below $T_c$, 
which is compatible with our findings. Clearly, to make more decisive conclusions, further experiments are needed.

\begin{acknowledgments}
We are benefited from useful discussions with A. Zawadowski.
This work
was supported by the Hungarian National Research Fund under grant numbers
OTKA T032162 and T037451.
\end{acknowledgments}

\bibliographystyle{apsrev}
\bibliography{eth}

\begin{thebibliography}{10}
\expandafter\ifx\csname bibnamefont\endcsname\relax
  \def\bibnamefont#1{#1}\fi
\expandafter\ifx\csname bibfnamefont\endcsname\relax
  \def\bibfnamefont#1{#1}\fi
\expandafter\ifx\csname url\endcsname\relax
  \def\url#1{\texttt{#1}}\fi
\expandafter\ifx\csname urlprefix\endcsname\relax\def\urlprefix{URL }\fi
\providecommand{\bibinfo}[2]{#2}
\providecommand{\eprint}[2][]{\url{#2}}

\bibitem{gruner}
\bibinfo{author}{\bibfnamefont{G.}~\bibnamefont{Gr\"uner}},
  \emph{\bibinfo{title}{Density waves in solids}}
  (\bibinfo{publisher}{Addison-Wesley}, \bibinfo{address}{Reading},
  \bibinfo{year}{1994}).

\bibitem{HR}
\bibinfo{author}{\bibfnamefont{B.~I.} \bibnamefont{Halperin}} \bibnamefont{and}
  \bibinfo{author}{\bibfnamefont{T.~M.} \bibnamefont{Rice}}, in
  \emph{\bibinfo{booktitle}{Solid State Physics}}, edited by
  \bibinfo{editor}{\bibfnamefont{F.}~\bibnamefont{Seitz}},
  \bibinfo{editor}{\bibfnamefont{D.}~\bibnamefont{Turnbull}}, \bibnamefont{and}
  \bibinfo{editor}{\bibfnamefont{H.}~\bibnamefont{Ehrenreich}}
  (\bibinfo{publisher}{Academic Press}, \bibinfo{address}{New York},
  \bibinfo{year}{1968}), vol.~\bibinfo{volume}{21}, p. \bibinfo{pages}{115}.

\bibitem{kopaev}
\bibinfo{author}{\bibfnamefont{L.~V.} \bibnamefont{Keldysh}} \bibnamefont{and}
  \bibinfo{author}{\bibfnamefont{Y.~V.} \bibnamefont{Kopaev}},
  \bibinfo{journal}{Fiz. Tverd. Tela} \textbf{\bibinfo{volume}{6}},
  \bibinfo{pages}{2791} (\bibinfo{year}{1964}).

\bibitem{Ners1}
\bibinfo{author}{\bibfnamefont{A.~A.} \bibnamefont{Nersesyan}}
  \bibnamefont{and} \bibinfo{author}{\bibfnamefont{G.~E.}
  \bibnamefont{Vachnadze}}, \bibinfo{journal}{J. Low T. Phys.}
  \textbf{\bibinfo{volume}{77}}, \bibinfo{pages}{293} (\bibinfo{year}{1989}).

\bibitem{Ners2}
\bibinfo{author}{\bibfnamefont{A.~A.} \bibnamefont{Nersesyan}},
  \bibinfo{author}{\bibfnamefont{G.~I.} \bibnamefont{Japaridze}},
  \bibnamefont{and} \bibinfo{author}{\bibfnamefont{I.~G.}
  \bibnamefont{Kimeridze}}, \bibinfo{journal}{J. Phys. Cond. Mat.}
  \textbf{\bibinfo{volume}{3}}, \bibinfo{pages}{3353} (\bibinfo{year}{1991}).

\bibitem{Ners3}
\bibinfo{author}{\bibfnamefont{A.~A.} \bibnamefont{Nersesyan}},
  \bibinfo{journal}{Phys. Lett. A} \textbf{\bibinfo{volume}{153}},
  \bibinfo{pages}{49} (\bibinfo{year}{1991}).

\bibitem{Schulz}
\bibinfo{author}{\bibfnamefont{H.~J.} \bibnamefont{Schulz}},
  \bibinfo{journal}{Phys. Rev. B} \textbf{\bibinfo{volume}{39}},
  \bibinfo{pages}{2940} (\bibinfo{year}{1989}).

\bibitem{GG}
\bibinfo{author}{\bibfnamefont{Z.}~\bibnamefont{Gul{\'a}csi}} \bibnamefont{and}
  \bibinfo{author}{\bibfnamefont{M.}~\bibnamefont{Gul{\'a}csi}},
  \bibinfo{journal}{Phys. Rev. B} \textbf{\bibinfo{volume}{36}},
  \bibinfo{pages}{699} (\bibinfo{year}{1987}).

\bibitem{marston}
\bibinfo{author}{\bibfnamefont{I.}~\bibnamefont{Affleck}} \bibnamefont{and}
  \bibinfo{author}{\bibfnamefont{J.~B.} \bibnamefont{Marston}},
  \bibinfo{journal}{Phys. Rev. B} \textbf{\bibinfo{volume}{37}},
  \bibinfo{pages}{3774} (\bibinfo{year}{1988}).

\bibitem{Ozaki}
\bibinfo{author}{\bibfnamefont{M.}~\bibnamefont{Ozaki}}, \bibinfo{journal}{Int.
  J. Quantum Chem.} \textbf{\bibinfo{volume}{42}}, \bibinfo{pages}{55}
  (\bibinfo{year}{1992}).

\bibitem{IO}
\bibinfo{author}{\bibfnamefont{H.}~\bibnamefont{Ikeda}} \bibnamefont{and}
  \bibinfo{author}{\bibfnamefont{Y.}~\bibnamefont{Ohashi}},
  \bibinfo{journal}{Phys. Rev. Lett.} \textbf{\bibinfo{volume}{81}},
  \bibinfo{pages}{3723} (\bibinfo{year}{1998}).

\bibitem{Sudip}
\bibinfo{author}{\bibfnamefont{C.}~\bibnamefont{Nayak}},
  \bibinfo{journal}{Phys. Pev. B} \textbf{\bibinfo{volume}{62}},
  \bibinfo{pages}{4880} (\bibinfo{year}{2000}).

\bibitem{benfatto}
\bibinfo{author}{\bibfnamefont{L.}~\bibnamefont{Benfatto}},
  \bibinfo{author}{\bibfnamefont{S.}~\bibnamefont{Caprara}}, \bibnamefont{and}
  \bibinfo{author}{\bibfnamefont{C.}~\bibnamefont{{Di Castro}}},
  \bibinfo{journal}{Eur. Phys. J. B} \textbf{\bibinfo{volume}{17}},
  \bibinfo{pages}{95} (\bibinfo{year}{2000}).

\bibitem{castroneto}
\bibinfo{author}{\bibfnamefont{A.~H.} \bibnamefont{Castro-Neto}},
  \bibinfo{journal}{Phys. Rev. Lett.} \textbf{\bibinfo{volume}{86}},
  \bibinfo{pages}{4382} (\bibinfo{year}{2001}).

\bibitem{nagycikk}
\bibinfo{author}{\bibfnamefont{B.}~\bibnamefont{D{\'o}ra}} \bibnamefont{and}
  \bibinfo{author}{\bibfnamefont{A.}~\bibnamefont{Virosztek}},
  \bibinfo{journal}{Eur. Phys. J. B} \textbf{\bibinfo{volume}{22}},
  \bibinfo{pages}{167} (\bibinfo{year}{2001}).

\bibitem{3dflux}
\bibinfo{author}{\bibfnamefont{D.~F.} \bibnamefont{Schroeter}}
  \bibnamefont{and} \bibinfo{author}{\bibfnamefont{S.}~\bibnamefont{Doniach}},
  \bibinfo{journal}{Phys. Rev. B} \textbf{\bibinfo{volume}{66}},
  \bibinfo{pages}{075120} (\bibinfo{year}{2002}).

\bibitem{nayak}
\bibinfo{author}{\bibfnamefont{S.}~\bibnamefont{Chakravarty}},
  \bibinfo{author}{\bibfnamefont{R.~B.} \bibnamefont{Laughlin}},
  \bibinfo{author}{\bibfnamefont{D.~K.} \bibnamefont{Morr}}, \bibnamefont{and}
  \bibinfo{author}{\bibfnamefont{C.}~\bibnamefont{Nayak}},
  \bibinfo{journal}{Phys. Rev. B} \textbf{\bibinfo{volume}{63}},
  \bibinfo{pages}{094503} (\bibinfo{year}{2001}).

\bibitem{carbotte}
\bibinfo{author}{\bibfnamefont{W.}~\bibnamefont{Kim}} \bibnamefont{and}
  \bibinfo{author}{\bibfnamefont{J.~P.} \bibnamefont{Carbotte}},
  \bibinfo{journal}{Phys. Rev. B} \textbf{\bibinfo{volume}{66}},
  \bibinfo{pages}{033104} (\bibinfo{year}{2002}).

\bibitem{scatter}
\bibinfo{author}{\bibfnamefont{B.}~\bibnamefont{D\'ora}},
  \bibinfo{author}{\bibfnamefont{A.}~\bibnamefont{Virosztek}},
  \bibnamefont{and} \bibinfo{author}{\bibfnamefont{K.}~\bibnamefont{Maki}},
  \bibinfo{journal}{Phys. Rev. B} \textbf{\bibinfo{volume}{66}},
  \bibinfo{pages}{115112} (\bibinfo{year}{2002}).

\bibitem{epl1}
\bibinfo{author}{\bibfnamefont{A.}~\bibnamefont{Virosztek}},
  \bibinfo{author}{\bibfnamefont{B.}~\bibnamefont{D{\'o}ra}}, \bibnamefont{and}
  \bibinfo{author}{\bibfnamefont{K.}~\bibnamefont{Maki}},
  \bibinfo{journal}{Europhys. Lett.} \textbf{\bibinfo{volume}{47}},
  \bibinfo{pages}{358} (\bibinfo{year}{1999}).

\bibitem{rapid}
\bibinfo{author}{\bibfnamefont{B.}~\bibnamefont{D\'ora}},
  \bibinfo{author}{\bibfnamefont{A.}~\bibnamefont{Virosztek}},
  \bibnamefont{and} \bibinfo{author}{\bibfnamefont{K.}~\bibnamefont{Maki}},
  \bibinfo{journal}{Phys. Rev. B} \textbf{\bibinfo{volume}{64}},
  \bibinfo{pages}{041101(R)} (\bibinfo{year}{2001}).

\bibitem{tesla}
\bibinfo{author}{\bibfnamefont{B.}~\bibnamefont{D\'ora}},
  \bibinfo{author}{\bibfnamefont{A.}~\bibnamefont{Virosztek}},
  \bibnamefont{and} \bibinfo{author}{\bibfnamefont{K.}~\bibnamefont{Maki}},
  \bibinfo{journal}{Phys. Rev. B} \textbf{\bibinfo{volume}{65}},
  \bibinfo{pages}{155119} (\bibinfo{year}{2002}).

\bibitem{imperfect}
\bibinfo{author}{\bibfnamefont{B.}~\bibnamefont{D{\'o}ra}},
  \bibinfo{author}{\bibfnamefont{K.}~\bibnamefont{Maki}}, \bibnamefont{and}
  \bibinfo{author}{\bibfnamefont{A.}~\bibnamefont{Virosztek}},
  \bibinfo{journal}{Phys. Rev. B} \textbf{\bibinfo{volume}{66}},
  \bibinfo{pages}{165116} (\bibinfo{year}{2002}).

\bibitem{impurd-wave}
\bibinfo{author}{\bibfnamefont{Y.}~\bibnamefont{Sun}} \bibnamefont{and}
  \bibinfo{author}{\bibfnamefont{K.}~\bibnamefont{Maki}},
  \bibinfo{journal}{Phys. Rev. B} \textbf{\bibinfo{volume}{51}},
  \bibinfo{pages}{6059} (\bibinfo{year}{1995}).

\bibitem{hotta}
\bibinfo{author}{\bibfnamefont{T.}~\bibnamefont{Hotta}}, \bibinfo{journal}{J.
  Phys. Soc. Jpn.} \textbf{\bibinfo{volume}{62}}, \bibinfo{pages}{274}
  (\bibinfo{year}{1993}).

\bibitem{epl2}
\bibinfo{author}{\bibfnamefont{B.}~\bibnamefont{D{\'o}ra}},
  \bibinfo{author}{\bibfnamefont{K.}~\bibnamefont{Maki}}, \bibnamefont{and}
  \bibinfo{author}{\bibfnamefont{A.}~\bibnamefont{Virosztek}},
  \bibinfo{note}{cond-mat/0012198}.

\bibitem{parks}
\bibinfo{author}{\bibfnamefont{K.}~\bibnamefont{Maki}}, in
  \emph{\bibinfo{booktitle}{Superconductivity}}, edited by
  \bibinfo{editor}{\bibfnamefont{R.~D.} \bibnamefont{Parks}}
  (\bibinfo{publisher}{Marcel Dekker}, \bibinfo{address}{New York},
  \bibinfo{year}{1969}).

\bibitem{klasszikus}
\bibinfo{author}{\bibfnamefont{A.~A.} \bibnamefont{Abrikosov}},
  \bibinfo{author}{\bibfnamefont{L.~P.} \bibnamefont{Gor'kov}},
  \bibnamefont{and} \bibinfo{author}{\bibfnamefont{I.~E.}
  \bibnamefont{Dzyaloshinski}}, \emph{\bibinfo{title}{Methods of Quantum Field
  Theory in Statistical Physics}} (\bibinfo{publisher}{Dover Publications},
  \bibinfo{address}{New York}, \bibinfo{year}{1963}).

\bibitem{pwave1}
\bibinfo{author}{\bibfnamefont{K.}~\bibnamefont{Maki}} \bibnamefont{and}
  \bibinfo{author}{\bibfnamefont{E.}~\bibnamefont{Puchkaryov}},
  \bibinfo{journal}{Europhys. Lett} \textbf{\bibinfo{volume}{45}},
  \bibinfo{pages}{263} (\bibinfo{year}{1999}).

\bibitem{pwave2}
\bibinfo{author}{\bibfnamefont{E.}~\bibnamefont{Puchkaryov}} \bibnamefont{and}
  \bibinfo{author}{\bibfnamefont{K.}~\bibnamefont{Maki}},
  \bibinfo{journal}{Eur. Phys. J. B} \textbf{\bibinfo{volume}{4}},
  \bibinfo{pages}{191} (\bibinfo{year}{1998}).

\bibitem{dressel}
\bibinfo{author}{\bibfnamefont{M.}~\bibnamefont{Dressel}},
  \bibinfo{author}{\bibfnamefont{N.}~\bibnamefont{Drichko}},
  \bibinfo{author}{\bibfnamefont{J.}~\bibnamefont{Schlueter}},
  \bibnamefont{and} \bibinfo{author}{\bibfnamefont{J.}~\bibnamefont{Merino}},
  \bibinfo{note}{cond-mat/0206074}.

\bibitem{greendw}
\bibinfo{author}{\bibfnamefont{K.}~\bibnamefont{Maki}}, \bibinfo{journal}{Phys.
  Rev. B} \textbf{\bibinfo{volume}{33}}, \bibinfo{pages}{4826}
  (\bibinfo{year}{1986}).

\bibitem{haran}
\bibinfo{author}{\bibfnamefont{G.}~\bibnamefont{Haran}} \bibnamefont{and}
  \bibinfo{author}{\bibfnamefont{A.~D.~S.} \bibnamefont{Nagi}},
  \bibinfo{journal}{Phys. Rev. B} \textbf{\bibinfo{volume}{45}},
  \bibinfo{pages}{15463} (\bibinfo{year}{1996}).

\bibitem{tutto1}
\bibinfo{author}{\bibfnamefont{I.}~\bibnamefont{T{\"u}tt{\H o}}}
  \bibnamefont{and}
  \bibinfo{author}{\bibfnamefont{A.}~\bibnamefont{Zawadowski}},
  \bibinfo{journal}{Phys. Rev. Lett.} \textbf{\bibinfo{volume}{60}},
  \bibinfo{pages}{1442} (\bibinfo{year}{1988}).

\bibitem{tutto2}
\bibinfo{author}{\bibfnamefont{I.}~\bibnamefont{T{\"u}tt{\H o}}}
  \bibnamefont{and}
  \bibinfo{author}{\bibfnamefont{A.}~\bibnamefont{Zawadowski}},
  \bibinfo{journal}{Phys. Rev. B} \textbf{\bibinfo{volume}{32}},
  \bibinfo{pages}{2449} (\bibinfo{year}{1985}).

\bibitem{hottacomment}
\bibinfo{author}{\bibfnamefont{T.}~\bibnamefont{Hotta}},
  \bibinfo{journal}{Phys. Rev. B} \textbf{\bibinfo{volume}{52}},
  \bibinfo{pages}{13041} (\bibinfo{year}{1995}).

\bibitem{szupravezetes}
\bibinfo{author}{\bibfnamefont{H.}~\bibnamefont{Won}} \bibnamefont{and}
  \bibinfo{author}{\bibfnamefont{K.}~\bibnamefont{Maki}}, in
  \emph{\bibinfo{booktitle}{Symmmetry and Pairing in Superconductors}}, edited
  by \bibinfo{editor}{\bibfnamefont{M.}~\bibnamefont{Ausloos}}
  \bibnamefont{and} \bibinfo{editor}{\bibfnamefont{S.}~\bibnamefont{Kruchinin}}
  (\bibinfo{publisher}{Kluwer}, \bibinfo{address}{Dordrecht},
  \bibinfo{year}{1999}).

\bibitem{rpa}
\bibinfo{author}{\bibfnamefont{B.}~\bibnamefont{D\'ora}} \bibnamefont{and}
  \bibinfo{author}{\bibfnamefont{A.}~\bibnamefont{Virosztek}},
  \bibinfo{journal}{Europhys. Lett} \textbf{\bibinfo{volume}{61}},
  \bibinfo{pages}{396} (\bibinfo{year}{2003}).

\bibitem{universal}
\bibinfo{author}{\bibfnamefont{Y.}~\bibnamefont{Sun}} \bibnamefont{and}
  \bibinfo{author}{\bibfnamefont{K.}~\bibnamefont{Maki}},
  \bibinfo{journal}{Europhys. Lett.} \textbf{\bibinfo{volume}{32}},
  \bibinfo{pages}{355} (\bibinfo{year}{1995}).

\bibitem{lee}
\bibinfo{author}{\bibfnamefont{P.~A.} \bibnamefont{Lee}},
  \bibinfo{journal}{Phys. Rev. Lett.} \textbf{\bibinfo{volume}{71}},
  \bibinfo{pages}{1887} (\bibinfo{year}{1993}).

\end{thebibliography}
\end{document}